\def \mathbi#1{\textbf{\em #1}}
\documentclass[3p,onecolumn]{elsarticle}
\usepackage{graphicx}
\usepackage{dcolumn}
\usepackage{pifont}
\usepackage{bm}
\usepackage{multirow}
\usepackage{amsmath}
\usepackage{float}
\usepackage{txfonts}

\journal{arXiv}

\begin{document}
\title{Self-consistent one-electron equation including exchange and correlation in terms of equivalent function and phase norm: calculation for ground and excited states in a unified way}

\author[Jong]{Chol Jong}
\author[Jong]{Byong-Il Ri}
\author[Yu]{Gwang-Dong Yu}
\author[Jong]{Son-Il Jo}
\author[Jong]{Sin-Hyok Jon}
\author[Jong]{Namchol Choe\corref{cor}}
\ead{cnc103@mail.dlut.edu.cn}

\cortext[cor]{Corresponding author}

\address[Jong]{ Faculty of Physics, Kim Chaek University of Technology, Kyogu-60, Yonggwang Street, Pyongyang, DPR Korea}
\address[Yu]{Faculty of Physics,  Kim Il Sung University, Ryongnam-Dong, Ryomyong Street, Pyongyang,DPR Korea}

\begin{abstract}

The ab initio calculation for many-electron systems sets up the extremely demanding tasks which involve the explicit expression of exchange and correlation, and the calculation for excited states in a way applicable to all stationary states. 
We present an alternative self-consistent one-electron equation which includes the exchange effect in an explicit way via the Slater determinant and the correlation effect in an original way according to the quantum principles of correlation.
To derive a one-electron equation including the exchange effect in an explicit way in terms of antisymmetric wavefunctions, we introduce a new concept called the equivalent function.
Moreover, to treat the electronic correlation in a first-principles way, we introduce another new concept referred to as the phase norm which specifies the mutual electron-approachable limit.
The derived equation becomes a self-consistent one-electron equation which satisfies the main requirements for ab initio calculations.
This equation enables us to calculate electronic states of  many-electron systems in a unified way commonly applicable to all stationary state problems, irrespective of the ground or excited states, without recourse to the approaches based on the Hartree-Fock method or the density functional theory. 
\end{abstract}


\maketitle

\section{\label{sec:intro} Introduction}

The main purpose of the many-particle theory in non-relativistic quantum mechanics is to study the properties of solutions of the Schr\"{o}dinger equation describing the characteristics of many-body interactions including the exchange and correlation effects. 
The key problems of ab initio calculations for many-particle systems are how to exclude the self-interaction, how to include the exchange and correlation effects and how to reduce the many-particle problem to a one-particle one.

By these criteria, we review the several approaches to ab initio calculations.
The complexity of many-electron problems is due to the interaction operator represented by two-particle variables.
It makes it impossible to use the variable separation method to solve the Schr\"{o}dinger equation for many-body systems.

The first approach to the many-electron problem proposed by Hartree and Slater came up with the idea of the introduction of one-electron wavefunction, the self-consistent field (SCF) approach and the applicability of the variational principle \cite{Hartree,Slater,Kohanoff}.
Although this approach is not realistic enough for many-body systems owing to no consideration of exchange and correlation, it provided the main route to many-electron approaches \cite{Kohanoff}. 
The Hartree method  does not reflect the antisymmetry of fermion and, in consequence, the description of many-particle systems in terms of it is incomplete.

The Hartree-Fock (HF) approximation conceived to follow Pauli's principle
makes it possible to take  into account electronic exchange in an explicit way by adopting the Slater determinant (SD) as a kind of antisymmetric wavefunction \cite{Slater,Fock}.
When neglecting interaction between electrons, from the point of view of mathematics, it is natural to choose as many-electron wavefunctions a single Slater determinant or the combination of the Slater determinants determined by means of a series of sophisticated techniques.
However, even in the presence of interaction, some aspects of the ground state of many-electron systems can be represented by a single Slater determinant \cite{Lipparini}. 
What is missing from this approximation is the electronic correlation.
The correlation problem which had not been dealt with in the early HF theory was treated to some extent, albeit with the aid of sophisticated methods in the post HF theory by using either more than one SD or other anti-symmetric wavefunctions of another type, and the post HF procedures addressing the electronic correlation have been considerably promoted  \cite{Jensen,Davidson}. 

In parallel to the approaches using single-electron orbitals, such as the Hartree and Hartree-Fock method, a  different method proposed by Thomas and Fermi takes the full electron density as the fundamental variable of the many-body problem.
The Hohenberg-Kohn theorem had underpinned in a rigorous way an alternative formalism, the density functional theory (DFT) using the electron density to describe the properties of the ground state (energy and electron density) of a many-particle system \cite{Hohenberg,Kohn}. 
As a self-consistent equation, the Kohn-Sham equation describes the properties of the ground state in the way that, formally, takes into account all many-body effects \cite{Kohanoff,Lipparini,Kohn}. 
This equation especially has  the decisive advantage of being framed as a one-electron equation \cite{Kohn}.
DFT has been applied most of all to electron systems, such as atoms, molecules, homogeneous solids, surfaces and interfaces, quantum wells and quantum dots etc, and has provided results in good accord with experiments \cite{Lipparini,McGUIRE,March}.

It should be noted that no one can doubt the successes of the modern Kohn-Sham approach to DFT, but this theory has bottlenecks such as the exclusion of self-interaction, analytical consideration of exchange and correlation.
Kohn-Sham's theory does not describe states of electron systems with the help of anti-symmetric wavefunctions. 
Therefore this equation cannot but employ the additional term relevant to both exchange and correlation more or less in a factitious way. 
In this connection, DFT has spawned a multitude of methods such as the local density approximation (LDA) \cite{Karasiev} and the generalized gradient approximation (GGA) \cite{Cafiero,Perdew5,Boese1,Boese2,Constantin} distinguished by the exchange and correlation calculations.
It should be noted that it is difficult to draw correlation functionals compatible with exact exchange functionals from a definite principle within the framework of the KS theory \cite{Cafiero}. 
This reason led to proposing several improved methods for DFT \cite{Cafiero,Constantin,Becke1,Becke2,Becke3,Sharma,Perdew1,Perdew3}.
One of the methods which had been explored to fix this problem makes use of exact exchange in the KS exchange-correlation calculation \cite{Cafiero,Gusarov}. 
Although there has been a great  advance in finding exact exchange functionals and correlation functionals compatible with it, the exchange-correlation problem also remains as a main challenge to reaching the goal of the ab initio theory. 
Such a situation of the exchange-correlation research in DFT shows that the proposed approaches are still far from completing the theory of ab initio calculations.

The present status of the excited-state quantum-chemical methodology can be assessed in terms of their reliability and common applicability \cite{Gonzalez}. 
The most commonly employed time-dependent-density functional theory (TD-DFT) methods can provide a general description of the excited state but not always guarantee high-accuracy description \cite{Stratmann,Kilina,Gross,Jamorski}. 
The configuration interaction-singles (CIS)-derived methods \cite{Dreuw1,Dreuw2,Roos} can be applicable not to any case. 
On the contrary, the linear response approaches find a broad range of application although the former approaches might fail to describe complex situations. 
The coupled-cluster approaches give a very accurate account of many types of excited states, provided that triple excitations are included and some conditions relevant to equilibrium geometries and multiconfigurational property of ground-state are fulfilled \cite{Nooijen1,Nooijen2,Koch}.
Many efforts have been made to obtain accurate molecular properties using the response theory with propagator approaches \cite{Falden,Oddershede,Packer}.
The recent survey of excited states indicates that a large number of studies use the TD-DFT methodology \cite{Gonzalez}. 

In this work, we aim to reach the goal of reducing the problem of many-electron systems to a one-electron approximation including the exchange-correlation in an explicit way.  
Meanwhile, we show that using the one-electron equation obtained offers the possibility of resolving problems of stationary states including ground and excited states in an identical way.

The remaining paper is organized in the following manner.
In Section 2 we deduce a one-electron equation including exchange effect in terms of the conceived equivalent function.
In Section 3 we deal with the electronic correlation problem in terms of the newly introduced phase norm.
The Section 4 is devoted to the calculation for ground and excited states in a unified way via the one-electron equation.
In Section 5 the summary and discussion are given. 
The paper is concluded in Section 6.

\section{One-electron equation including exchange effect in terms of antisymmetric wavefunction}

Our goal is to obtain a one-electron equation with respect to single-electron orbitals including the exchange in an original way with the help of antisymmetric wavefunction.
We start deducing a one-electron equation from the variational principle of total energy functional with SD. 
Therefore the most part of our deduction follows HF approach.
However our last manipulation results in a single-electron eigenvalue equation different from HF equation.

To begin with, we review the consistency of the Hartree-Fock equation.

The Hartree-Fock equation can be written in a compact shape as 
\begin{eqnarray}\label{eq:10a}
\hat{H}_1 \psi_i\left(\mathbf{r}_1\right)+ E^{\left(int\right)}_i\left(\mathbf{r}_1\right)\psi_i\left(\mathbf{r}_1\right)+E^{\left(exc\right)}_i\left(\mathbf{r}_1\right)\psi_i\left(\mathbf{r}_1\right)=\epsilon_i\psi_i\left(\mathbf{r}_1\right),
\end{eqnarray}   
where
\begin{eqnarray}
E^{\left(exc\right)}_i\left(\mathbf{r}_1\right)=\left\{-\sum_{j \neq i}  \left[\delta \left(\mathit{m}_{si}, m_{sj}\right) \times\int \psi_j^\ast\left(\mathbf{r}_2\right) \hat{H}_{12}\psi_i\left(\mathbf{r}_2\right)\mathrm{d}\tau_2\right] \frac{\psi_j\left(\mathbf{r}_1\right)}{\psi_i\left(\mathbf{r}_1\right)}\right\}
\end{eqnarray}  
and 
\begin{eqnarray}
E^{\left(int\right)}_i\left(\mathbf{r}_1\right)=\sum_{j \neq i}\left[\int \psi_j^\ast\left(\mathbf{r}_2\right) \hat{H}_{12}\psi_j\left(\mathbf{r}_2\right)\mathrm{d}\tau_2\right].
\end{eqnarray}
Equation \ref{eq:10a} can be considered as the eigenvalue equation for the $i$th  one-electron state.
The eigenvalue equation varies with $i$. 
Therefore every eigenvalue equation does not give the same eigenfunctions as those the others yield, and thus the eigenfunctions of a given eigenvalue equation are not orthogonal to those of the other eigenvalue equations.

Obviously, a one-electron equation should be represented as a single equation satisfying the following relation, 

\begin{subequations}
 \begin{gather}
 \label{eq:34}
\hat{H}\psi\left(\mathbi{r}\right)=\epsilon \psi\left(\mathbi{r}\right),\\
 \label{eq:35}
\varepsilon=\sum_i \epsilon_i,
\end{gather}     
\end{subequations}
where $\varepsilon$ is the total energy of an electron system  and $\epsilon_i$ is the $i$th eigenvalue of the eigenvalue equation \ref{eq:34}.
Since one-electron wavefunctions constitute an orthogonal basis, there necessarily exists a single eigenvalue equation corresponding to an orthogonal basis.
The case should be the same for HF method.
It is obvious that according to this criterion, the Kohn-Sham equation is a one-electron equation whereas the Hartree equation and the HF equation are otherwise. 

Now, we examine whether the HF equation gives consistent eigenstates.
Generally, the system of HF equation can be written as
\begin{eqnarray}
 \label{eq:37}
\hat{H}_i\psi_i\left(\mathbi{r}\right)=\epsilon_i \psi_i\left(\mathbi{r}\right).
\end{eqnarray} 
If these eigenvalue equations are consistent, Eqs. \ref{eq:37} should yield mutually orthogonal eigenfunctions.
If the eigenfunctions are supposed to be orthogonal, by multiplying the both sides of Eq. \ref{eq:37} by $\psi_j^*$ orthogonal to $\psi_i$ and by integrating them, we get
\begin{eqnarray}
 \label{eq:38}
\int\psi_j^*\hat{H}_i\psi_i\left(\mathbi{r}\right)\mathrm{d}\tau=\int\psi_j^*\epsilon_i \psi_i\left(\mathbi{r}\right)\mathrm{d}\tau=0.
\end{eqnarray} 
Likewise, for $\hat{H}_j \neq \hat{H}_i $ we obtain
\begin{eqnarray}
 \label{eq:39}
\int\psi_i^*\hat{H}_j\psi_j\left(\mathbi{r}\right)\mathrm{d}\tau=\int\psi_i^*\epsilon_j \psi_j\left(\mathbi{r}\right)\mathrm{d}\tau=0.
\end{eqnarray} 
Considering the Hermitian property of $\hat H_i$ and $\hat H_j$, and taking the complex conjugate of Eq. \ref {eq:39}, we can recast Eq. \ref{eq:39} as
\begin{eqnarray}
\label{eq:40}
\int\psi_j^*\hat{H}_j\psi_i\left(\mathbi{r}\right)\mathrm{d}\tau=0.
\end{eqnarray} 
It follows from Eq. \ref{eq:38} and Eq. \ref{eq:40} that $\hat H_i=\hat H_j$.
Therefore the assumption about the orthogonality of $\psi_i$ and $\psi_j$ ends up with contradiction.
This contradiction is ascribed to the inconsistency of the Hartree and HF equation.
This inconsistency consists in considering one-electron wavefunctions $\psi_i$ as independent ones.
In fact, wavefunctions $\psi_i$ should be subject to the orthonormality constraint so that they cannot be completely independent.
A single eigenvalue equation such as Eq. \ref{eq:34} ensures the constraint of orthonormality but the system of eigenvalue equations such as HF equation cannot.  
As a consequence, HF equation cannot describe true electronic states fulfilling the orthonomality.
To prevent such an inconsistency, it is indispensible to construct a one-electron equation in the form of Eq. \ref{eq:34}.

It should be considered that electrons are not distinguished, being identical particles and, consequently,  prove to be in the same quantum-mechanical state through exchange.   
Therefore we require that the eigenvalue equation for an electron system should be determined so that it can yield a definite set of eigenfunctions $\psi_i$ describing not an electron but the electron system so as to agree with the identity principle.

It is obvious that the variation of an eigenfunction is not independent of the others and, consequently, the variation of the total energy functional in terms of SD, 
$
\Phi= \langle H \rangle -\epsilon \langle \Psi | \Psi \rangle
$
should be manipulated as a whole without separation.
Namely,
\begin{eqnarray}
\label{eq:11}
\delta\Phi=\sum_i\delta_i\Phi=0.
\end{eqnarray}                    
In other words, the variation for a many-electron system cannot be separated into individual variations with respect to one-electron wavefunctions.

Now, it is necessary to find a single eigenvalue equation from the variational equation \ref{eq:11} using a special method enabling the reduction of many-body problems to one-body problems. 
This is the mathematical requirement that originates from the identity principle.
To resolve this problem, we consider in a different way the variation with respect to
$\left\{\psi_i\left(\mathbf{r}_1\right) |i=0, 1, 2, \cdots, N \right\}$.
Obviously, it is impossible to perform arbitrary independent variations with respect to 
$\left\{\psi_i\left(\mathbf{r}_1\right) |i=0, 1, 2, \cdots, N \right\}$
and admissible vatiations are limited to the unitary transformation. 

By performing the variation, $
\left\{\delta\psi_i\left(\mathbf{r}_1\right), \delta\psi_i^\ast\left(\mathbf{r}_1\right) |i=0, 1, 2, \cdots, N \right\}$,
we get the following equation:
\begin{eqnarray}
\label{eq:14}
\sum_i\delta_i\Phi=\sum_i\int\delta\psi^\ast_i\left(\mathbf{r}_1\right)\left\{ \hat{H}_1+E^{(int)}_i \left(\mathbf{r}_1\right)+E^{(exc)}_i \left(\mathbf{r}_1\right)-\epsilon_i \right\}\psi_i\left(\mathbf{r}_1\right)\mathrm{d}\tau_1\nonumber\\
+c.c=0.
\end{eqnarray}  
To deduce a one-electron equation describing electron systems, we introduce a definite equivalent function $\psi\left(\mathbf{r}_1\right)$  and $\varepsilon$ determined by 
$\left\{\psi_i\left(\mathbf{r}_1\right) |i=0, 1, 2, \cdots, N \right\}$
and $\left\{\epsilon_i |i=0, 1, 2, \cdots, N \right\}$.

Next, it is assumed that for the equivalent function $\psi\left(\mathbf{r}_1\right)$ and equivalent energy $\varepsilon$, the following relations: 
\begin{subequations}
 \begin{gather}
\sum_{i=1}^N\delta\psi^\ast_i\left(\mathbf{r}_1\right)\hat{H}_1\psi_i\left(\mathbf{r}_1\right)=\delta\psi^\ast\left(\mathbf{r}_1\right)\left[N\hat{H}_1\right]\psi\left(\mathbf{r}_1\right), \label{eq:17}\\
\sum_{i=1}^N\delta\psi^\ast_i\left(\mathbf{r}_1\right)E^{(int)}_i\psi_i\left(\mathbf{r}_1\right)=\delta\psi^\ast\left(\mathbf{r}_1\right)\left[\sum_{i=1}^NE^{(int)}_i\right] \psi\left(\mathbf{r}_1\right), \label{eq:18}\\
\sum_{i=1}^N\delta\psi^\ast_i\left(\mathbf{r}_1\right)E^{(exc)}_i \cdot \psi_i\left(\mathbf{r}_1\right)=\delta\psi^\ast\left(\mathbf{r}_1\right)\left[\sum_{i=1}^NE^{(exc)}_i\right]\psi\left(\mathbf{r}_1\right), \label{eq:19}\\
\sum_{i=1}^N\delta\psi^\ast_i\left(\mathbf{r}_1\right)\epsilon_i\psi_i\left(\mathbf{r}_1\right)=\delta\psi^\ast\left(\mathbf{r}_1\right)\left[N\varepsilon\right] \psi\left(\mathbf{r}_1\right) \label{eq:20}
\end{gather}     
\end{subequations}
should hold.
Then $\left\{\psi_i\left(\mathbf{r}_1\right) |i=0, 1, 2, \cdots, N \right\}$ in the operator terms are the eigenfunctions determined by the eigenvalue equation for equivalent function  $\psi\left(\mathbf{r}_1\right)$.

As a result of the above assumed relation, the eigenvalue equation for the equivalent function satisfying the extremum condition for total energy functional can be represented as  
\begin{eqnarray}
 \label{eq:21}
\int\delta\psi^\ast\left(\mathbf{r}_1\right)\left\{N \hat{H}_1+\sum_iE^{(int)}_i \left(\mathbf{r}_1\right)+\sum_iE^{(exc)}_i \left(\mathbf{r}_1\right)-N\varepsilon \right\}\psi\left(\mathbf{r}_1\right)\mathrm{d}\tau_1=0.
\end{eqnarray}                                                                                        
Here the exchange operator in the form of integral is written as
\begin{align}
 \label{eq:22}
E^{(exc)}_i \left(\mathbf{r}_1\right) &=-\sum_{j \neq i} \left[\delta\left(m_{s_i}, m_{s_j}\right) \int \psi^\ast_j\left(\mathbf{r}_2\right)\hat{H}_{12}\psi_i\left(\mathbf{r}_2\right)\mathrm{d}\tau_2 \right] \frac{\psi_j\left(\mathbf{r}_1\right)}{\psi_i\left(\mathbf{r}_1\right)} \nonumber\\
&=-\sum_{j \neq i} \left[\delta\left(m_{s_i}, m_{s_j}\right) \int \psi^\ast_j\left(\mathbf{r}_2\right)\hat{H}_{12}\psi_i\left(\mathbf{r}_2\right)\mathrm{d}\tau_2 \right] \frac{\psi^\ast_i\left(\mathbf{r}_1\right)\psi_j\left(\mathbf{r}_1\right)}{\psi^\ast_i\left(\mathbf{r}_1\right)\psi_i\left(\mathbf{r}_1\right)} \nonumber\\
&=-\sum_{j \neq i} \left[\delta\left(m_{s_i}, m_{s_j}\right) \int \rho_{ji}\left(\mathbf{r}_2\right)\hat{H}_{12}\mathrm{d}\tau_2\right] \frac{\rho_{ij}\left(\mathbf{r}_1\right)}{\rho_i\left(\mathbf{r}_1\right)} ,
\end{align}   
the direct-interaction operator in the form of integral, as
\begin{eqnarray}
 \label{eq:23} 
\sum_{i=1}^{N}E^{(int)}_i \left(\mathbf{r}_1\right)&=\left(N-1\right)\sum_{j=1}^{N}\int  \psi^\ast_j\left(\mathbf{r}_2\right)\hat{H}_{12}\psi_j\left(\mathbf{r}_2\right)\mathrm{d}\tau_2 \nonumber \\
&=\left(N-1\right)\sum_{j=1}^{N}\int  \rho_j\left(\mathbf{r}_2\right)\hat{H}_{12}\tau_2 .
\end{eqnarray}   

Thus,  we derive a single eigenvalue equation for \textit{N}-electron system:   
\begin{eqnarray}
 \label{eq:25}
N \hat{H}_1\psi\left(\mathbf{r}_1\right)+\sum_iE^{(int)}_i \left(\mathbf{r}_1\right)\psi\left(\mathbf{r}_1\right)+\sum_iE^{(exc)}_i \left(\mathbf{r}_1\right)\psi\left(\mathbf{r}_1\right)=N\varepsilon \psi\left(\mathbf{r}_1\right).
\end{eqnarray}  
A further arrangement gives
\begin{align}
 \label{eq:26}
 & \hat{H}_1\psi\left(\mathbf{r}_1\right)+\frac{N-1}{N}\left[\sum^N_{j=1}\int\psi^\ast_j\left(\mathbf{r}_2\right) \hat{H}_{12}\psi_j\left(\mathbf{r}_2\right)\mathrm{d}\tau_2\right]\psi\left(\mathbf{r}_1\right)\nonumber\\ &+\frac{1}{N}\left[\sum_{i=1}^{N}E^{(exc)}_i \left(\mathbf{r}_1\right)\right]\psi\left(\mathbf{r}_1\right)=\varepsilon \psi\left(\mathbf{r}_1\right).
\end{align}  

In the physical sense, Eq. \ref{eq:26} can be inferred as follows.
Equation \ref{eq:10a} can be considered as eigenvalue equations slightly different from the exact one-electron equation.
Therefore \textit{N} one-electron equations are represented as
\begin{eqnarray}\label{eq:26a}
\hat{H}_1 \psi\left(\mathbf{r}_1\right)+ E^{(int)}_i\left(\mathbf{r}_1\right)\psi\left(\mathbf{r}_1\right)+E^{(exc)}_i\left(\mathbf{r}_1\right)\psi\left(\mathbf{r}_1\right)=\epsilon\psi\left(\mathbf{r}_1\right).
\end{eqnarray}   
Equation \ref{eq:26a} can be recast shortly as
\begin{eqnarray}\label{eq:26c}
\mathcal{H}_i\psi\left(\mathbf{r}_1\right)=\epsilon\psi\left(\mathbf{r}_1\right).
\end{eqnarray}   
In this case, the eigenvalue equations vary with $E^{(int)}_i\left(\mathbf{r}_1\right)$ and $E^{(exc)}_i \left(\mathbf{r}_1\right)$, i.e. $\mathcal{H}_i$ .
Therefore the eigenvalue equations produce different systems of eigenfunctions.
As a result, every eigenvalue equation can be thought of as deviating from the exact one-electron equation for a given \textit{N}-electron system.
To absorb the differences, we introduce the mean operator given by $\mathcal{H}_i$.
As a consequence, we obtain the one-electron equation \ref{eq:26}.

Another interpretation on Eq. \ref{eq:26}  is available.
Equation \ref{eq:26a} stands for the $i$th one-electron state.
Meanwhile, it should be considered that every electron of a \textit{N}-electron system passes through all one-electron states with an equal probability by means of exchange.
On this account, all electrons of a many-electron system are identical.
Therefore the wave equation for many-electron systems should be represented by a one-electron equation, and then the Hamiltonian of an electron should be defined by the average Hamiltonian operator,  $\sum_i\mathcal{H}_i/N$ because $\mathcal{H}_i$ is subject to the law of so-called equipartition of probability.
By this consideration, we again get Eq. \ref{eq:26}.

By using the one-electron wavefunctions given by the eigenvalue equation \ref{eq:26}, we can calculate the total energy of a many-electron system on

\begin{align}
E&=\sum^N_{k=1}\int\psi^\ast_k\left(\mathbf{r}_1\right)\left\{\hat{H}_1+\frac{N-1}{N}\left[\sum^N_{j=1}\int\psi^\ast_j\left(\mathbf{r}_2\right) \hat{H}_{12}\psi_j\left(\mathbf{r}_2\right)\mathrm{d}\tau_2\right]  
\right.\nonumber\\
&\left.+\frac{1}{N}\left[\sum_{i=1}^{N}E^{(exc)}_i\left(\mathbf{r}_1\right)\right]\right\}\psi_k\left(\mathbf{r}_1\right)\mathrm{d}\tau_1 \nonumber\\
&=\sum^N_{k=1}\int\psi^\ast_k\left(\mathbf{r}_1\right)\varepsilon_k\psi_k\left(\mathbf{r}_1\right)\mathbf{d}\tau_1 =\sum^N_{k=1}\varepsilon_k~.
\end{align}

In this manner, the eigenvalue equation \ref{eq:26} describes the electronic states of many-electron systems with the help of one-electron wavefunctions.

\section{Consideration of electronic correlation in terms of phase norm}

Generally, it is considered that the correlation in a wavefunction is represented as the difference between the exact wavefunction and the HF wavefunction.
Is it possible to find the exact wavefunction without the knowledge of the nature of correlation?
It is impossible to do so.
To resolve the correlation problem, before everything, it is necessary to understand the quantum nature of the electronic correlation.
The nature of correlation has come into question from early time \cite{Chapman,Gouy,Debye} but the present physics of correlation is too uncertain to draw firm conclusions.
The correlation in a wavefuction is not ascribed only to the Coulomb interaction. 

Obviously, the electronic correlation is attributed to the limitation on the configuration of electrons imposed by the reason that electrons are impossible to get infinitely close to one another due to quantum   cause.
It is considered that from the point of view of quantum mechanics, the correlation between electrons is attributed to the uncertainty relation and Pauli's principle.

A variety of methods for taking account of the electronic correlation have been proposed but the present situation of the correlation theory seems not to embody the nature of electronic correlation.
Since  the electronic correlation originally is relevant to the approachable limit of particles, it is consistent to take account of it through the integral of the operator term by restricting the approach of particles in a quantum-mechanical way.        
This problem can be solved by inserting a correlation-hole function in terms of the phase norm into the integral of two-body operator term in the obtained one-electron equation without employing any additional term relevant to correlation.
The correlation-hole function gives the value 1 within the admissible range of electron approach and the value 0 out of the range.    
The key problem is from what principle the approachable limit of electrons should be specified.

We use phase space to specify the approachable limit of electrons.
If two electrons are involved in the same volume as $h^3$ of phase space, it is proved that they are in the same quantum-mechanical state. 
Consequently, such an approach is prohibited by the uncertainty principle and Pauli's principle.     
The main point of the present method consists in multiplying the integrand of the two-body operator term of Eq. \ref{eq:26} by a correlation-hole function determined by evaluating the phase norm for two electrons. 
Using the correlation-hole function $\theta \left(\mathbf{r}_1, \mathbf{r}_2\right)$, we can rewrite the one-electron equation as 
\begin{align}
\label{eq:28}
 & \hat{H}_1\psi\left(\mathbf{r}_1\right)+\frac{N-1}{N}\left[\sum^N_{j=1}\int\psi^\ast_j\left(\mathbf{r}_2\right) \hat{H}_{12}\psi_j\left(\mathbf{r}_2\right)\theta \left(\mathbf{r}_1, \mathbf{r}_2\right)\mathrm{d}\tau_2\right]\psi\left(\mathbf{r}_1\right) \nonumber\\
& +\frac{1}{N}\left\{-\sum_{j \neq i} \left[\delta\left(m_{s_i}, m_{s_j}\right) \int \psi^\ast_j\left(\mathbf{r}_2\right)\hat{H}_{12}\psi_i\left(\mathbf{r}_2\right)\theta \left(\mathbf{r}_1, \mathbf{r}_2\right)\mathrm{d}\tau_2 \right]\frac{\psi_j\left(\mathbf{r}_1\right)}{\psi_i\left(\mathbf{r}_1\right)} \right\}\psi\left(\mathbf{r}_1\right)
\nonumber\\
&=\varepsilon \psi\left(\mathbf{r}_1\right).
\end{align}  

To determine the correlation-hole function, we should evaluate whether the states of two electrons are identical or not.

For this purpose, we define the distance between two states in phase space. 
First, for two particles we determine the magnitude of momentum of a particle with respect to the other and the distance from the particle to the other as
\begin{gather}\label{eq:29}
\Delta p=\left|\mathbi{p}_1-\mathbi{p}_2\right|,\\
\Delta r=\left|\mathbi{r}_1-\mathbi{r}_2\right|.
\end{gather}
Next, we define the phase norm as  
\begin{equation}\label{eq:30a}
\varrho =\Delta r\cdot\Delta p.
\end{equation}
Equation \ref{eq:30a} determines the distance between the states of two particles in phase space.
If the phase norm is smaller than the Plank constant $h$, it means that two particles are in the same quantum-mechanical state.
Such a situation is prohibited by the quantum principles.
Therefore it is natural to consider that the correlation-hole function in terms of the phase norm should be the following two-value function:

\begin{equation}
\theta=
\begin{cases}
0 & ; \varrho < 2\pi\hbar \\
1 & ; \varrho \geq 2\pi\hbar~~~.
\end{cases} 
\end{equation}
In practice, we can introduce an appropriate fitting constant $\alpha$ so as to have the following relation:
\begin{equation}
\theta=
\begin{cases}
0 & ; \alpha\varrho < 2\pi\hbar \\
1 & ; \alpha\varrho \geq 2\pi\hbar~~~.
\end{cases} 
\end{equation}
This definition reflects the requirement that the approach of particles should be admitted under the condition that the volume element in phase space assigned to every particle is $h^3$.

Next, we represent the momentum of a particle with the help of  the wavefunction. 
Taking into consideration the relation between the probability density and probability current density, we have 
\begin{eqnarray}
 \label{eq:31}
\mathbi{j}=\rho \mathbi{v}=\frac{\rho \mathbi{p}}{m}.
\end{eqnarray}     
where $\mathbi{v}$ is the velocity of an electron and $m$, the mass of the electron.
Accordingly, we get the expression for momentum
\begin{eqnarray}
 \label{eq:32}
\mathbi{p}=\frac{m \mathbi{j}}{\rho}.
\end{eqnarray}
Using the $\mathbi{i}$ wavefunction, we represent the momentum of an electron as
\begin{align}
 \label{eq:33}
\mathbi{p}_i\left(\mathbi{r}\right)&=\frac{i\hbar}{2\psi_i^\ast\psi_i}\left(\psi_i\nabla \psi_i^\ast-\psi_i^\ast\nabla \psi_i\right)\nonumber\\
&=\frac{i\hbar}{2}\left(\frac{1}{\psi_i^\ast}\nabla \psi_i^\ast-\frac{1}{\psi_i}\nabla \psi_i\right)=\hbar \mathrm{Im}\left(\frac{1}{\psi_i}\nabla \psi_i\right),
\end{align}
where $\mathbi{r}$ is the position variable of an electron, $i$ denotes the $i$th electronic state and $\mathrm{Im}$, the imaginary part of a complex number.

According to Eq. \ref{eq:33}, it is obvious that the phase norm comprises the electron density and its gradient. 
This fact implies that the phase norm method is closely related to LDA and GGA.
In this way, we have introduced a kind of correlation-hole function in terms of the phase norm to take into consideration the correlation effect in a first-principles way that embodies the quantum nature of correlation. 

\section{\label{sec:Excited}Calculation for ground and excited states in a unified way}

In general, a complete description of electronic excitation covers the knowledge about the ground state, the excited states and the way that couples the external perturbation to the electronic states of the system under study.
It is commonly accepted that the basic problems of the calculation for excited states are to derive acceptable approximations for the total energy functional of excited states and to find suitable variation principle for these states.

Our aim is to elucidate excited states of an electron system in a unified way applicable to both ground and excited states in terms of the one-electron equation unlike the linear response theory and DFT.
The reason for this is that the linear response theory and DFT originally involve a definite approximation since they are affected by such limitations as given by introducing factitiously the time-dependent relation even into stationary-state problems.

In the sense of the independent-particle model (IPM), the electronic excited state is the electronic state with inner unoccupied states.
Within the framework of the theory of one-electron equation, a many-electron state is expressed by one-electron wavefunctions constituting a vector, briefly, the state vector.
Since the total energy functional depends on this state vector, the dimension of state is equal to the number of electrons of the system. 
Therefore the excited-state problem of many-electron systems can be defined as the problem to determine all state vectors except for ground state vector giving extreme values of the total energy functional defined in the function space with same dimension as the number of electrons. 

We consider that the total energy functional, in principle, comprises all information about the ground and excited states because they  are both the stationary state. Therefore in essence the problem of excited-state is not distinguished from one of ground state.

In IPM, excited states are easily calculated.
As excited state, the state with an unoccupied level can be obtained in such a way that  with \textit{N} electrons, one fills \textit{N}-1 one-electron states below the Fermi level and a one-electron state above the Fermi level.
Such an understanding of excited states can be extended to the excited-state problem with arbitrary unoccupied states. 
Similarly, the state with $f$ unoccupied levels can be obtained in such a way that  with \textit{N} electrons, one fills $N-f$ one-electron states below the Fermi level and $f$ states above the Fermi level.

Let us consider the energy of an excited state with two unoccupied states.
For convenience, we denote the excited state of the many-electron system by 
\begin{align}
\vert i^{-1}j^{-1}; mn\rangle,
\end{align}
where $i^{-1}, j^{-1}$ denote unoccupied one-electron states below the Fermi level and $mn$, occupied one-electron states above the Fermi level.
Then since the system is a system of non-interacting particles, the energy of the system is represented as 
 \begin{align}
 E_0+\varepsilon_m+\varepsilon_n-\varepsilon_i-\varepsilon_j
\end{align}
where $E_0$ is the energy of the ground state of \textit{N}- electron system without unoccupied states and $\varepsilon_m, \varepsilon_n, \varepsilon_i, \varepsilon_j$,  the energies of individual levels corresponding to $m, n, i^{-1}, j^{-1}$.
Since IPM neglects interaction between particles, one-electron states and the corresponding energies do not depend on other occupied or unoccupied states.
However, for interacting systems the case is different although one uses the one-electron equation.
As the one-electron state vector varies with occupied states, so  does the corresponding energy.
On this account, for interacting systems excited states are not possible to be determined directly from the information on a ground state.

To the best of our knowledge, the one-electron equation opens up the possibility of resolving all the stationary state problems, whether ground or excited states, by means of a unified method.
Within the theory of one-electron equation, the state vector is represented in function space of the same dimension as the number of particles of the system.
The number of electrons of a given system defines the dimensionality of the function space standing for the many-electron system.
In general, the total energy functional has a series of extreme points in function space.
Obviously, these extreme points correspond to excited states of a many-electron system.

In our one-electron model, the problem of excited states is to be resolved in such a way that one takes a definite one-electron state vector in the vicinity of an extreme value of the total energy functional as the initial point and then makes this point converge at the exact extreme point by performing the self-consistent calculation.
In this case the self-consistent calculation can be considered as playing the role of unitary transformation which transforms an approximate  state vector chosen in the vicinity of an extreme value into the state vector for the exact extreme value.

The key problem is how to choose the approximate state vector guaranteeing the convergence in the vicinity of the exact extreme value.
The one-electron equation for interacting many-electron systems provides a possibility of determining excited states easily in a consistent way.
This calculation for excited states by the use of one-electron equation begins with searching for an approximate extreme point of a state vector in the vicinity of the exact extreme point of  the state vector.
We denote an excited state by $
\vert i^{-1}j^{-1}\cdots; mn\cdots \rangle$, where  $ i^{-1}j^{-1}\cdots$ denotes unoccupied states and $mn\cdots$,  occupied states.
Such an understanding of excited states is an approximation in terms of the intuitive picture of the one-electron state for non-interacting electron systems.
For a given excited state, we first consider the ground state of the system where all  unoccupied states below the highest level are supposed to be filled with electrons.
In this case, the electron system increases in the number of electrons by the number of unoccupied levels and the state vector is represented as 
\begin{align}
\vert ij\cdots mn\cdots \rangle.
\end{align}
The ground state of this many-electron system can be determined by solving the one-electron equation \ref{eq:28}.
 
One-electron wavefunctions, $\vert i \rangle$, $\vert j \rangle \cdots$ $\vert m\rangle$, $\vert n\rangle\cdots$constitutes the state vector,
$\vert ij\cdots mn\cdots \rangle$.
Meanwhile, these one-electron wavefunctions depend on occupied states.
Therefore the one-electron wavefunctions of the ground state are different from those of the excited state to some extent.
It is well-grounded to say that these corresponding one-electron states constituting ground and excited states approximate to each other.
In fact, the effect of unoccupied or occupied states on these states can be considered as a perturbation.
Therefore we can choose as an initial point for searching for excited states the combination of the state vectors of non-interacting system and ground state vectors of interacting system where unoccupied states are fully filled.
This combination can be represented as
 \begin{eqnarray}
\psi_i^{0}=\alpha_i \psi_i^{(non-int)}+\left( 1-\alpha_i\right) \psi_i^{(int)},
\end{eqnarray}
where $\psi_i^{0}$ is the $i$th one-electron wavefunction of the initial state vector, $\psi_i^{(non-int)}$ the one-electron wavefunction of the ground state vector in non-interacting case and  $\psi_i^{(int)}$ the one-electron wavefunction of the ground state vector in interacting case.
The range of $\alpha_i$ is the interval $[0, 1]$.
Therefore the initial state vector $\psi_i^{0}$ approaches $\psi_i^{(non-int)}$, as $\alpha_i$ tends to 1.
Conversely, $\psi_i^{0}$ approaches $\psi_i^{(int)}$, as $\alpha_i$ tends to 0.
By taking a suitable $\alpha_i$, one can choose a proper initial state vector guaranteeing the best convergence.

Next, the initial state vector should be forced to approach a definite state vector giving an exact extreme of the total energy functional by means of the unitary transformation.
This transformation can be implemented by the self-consistent calculation in terms of the one-electron equation. 
The previous-step state vector is transformed into a new state vector with orthonormal one-electron wavefunctions with the help of the one-electron equation.
This iterative procedure is repeated until the difference
between the state vectors obtained in two successive iterations falls below some predefined accuracy criterion.
The self-consistent calculation via the one-electron equation is identified with the unitary transformation, since it transforms an orthonormal basis into another.
Therefore the self-consistent calculation in terms of the one-electron equation becomes a process of convergence of a state vector at an extreme point giving the total energy functional.
In this way, it is possible to choose initial points for finding all excited states on the basis of the picture of one-electron state and to search for exact extreme points with the help of self-consistent calculation. 
In this context, the calculation for the ground and excited states cannot be distinguished.
The only difference is that the ground state corresponds to the lowest extremal value of the total energy functional, while the excited states coincide with the rest extremal values of the total energy functional.
Thus, the methodology for the calculation for all stationary states including both ground and excited states is achieved.

\section{\label{sec:Summary}Summary and discussion}

We summarize the main results.

First, we have obtained an alternative self-consistent one-electron equation for many-particle systems which describes the exchange effect in an explicit way. 
The deduction of this equation begins with antisymmetric wavefunction via SD.
To derive the one-electron equation, we introduced the so-called equivalent function.
The equivalent function bears the meaning analogous to the pseudo wavefunction.
This equation is based on single-electron orbitals instead of the density in DFT and has the feature of one-electron equation similar to the Kohn-Sham equation.

Second, we have shown that it is meaningful and useful to introduce the correlation-hole function in terms of the phase norm to consider the correlation effect in an explicit way.
The phase norm can be used to reveal the cause of pure quantum correlation.
As a consequence, this one-electron equation fulfills all the requirements for the exclusion of self-interaction, explicit expression of exchange and correlation, and reduction of many-electron equation to one-electron equation in a consistent way.

Third, we have demonstrated that it is possible to resolve the problems of the calculation for ground and excited states in a unified and self-consistent way.

The features of this approach can be understood as follows.

This approach employs single-electron orbitals.
As a result, the self-interaction is automatically excluded within the framework of this approach unlike the Kohn-Sham equation employing the electron density.
In order to formulate  the quantum many-particle theory, it is necessary to grasp what significance the concept of the one-electron state has.
Since quantum many-electron systems are the systems of identical particles, electrons are not distinguished.
Therefore the one-electron state is feasible and indispensable for describing many-electron systems.
The obtained equation does not describe the real one-electron states in the presence of electron interaction.
In the exact sense, this equation can be considered to describe states of quasi-particles or electronic plasmons standing for a many-electron system.
We consider that the real one-electron state loses its meaning because electrons are not distinguished from one another due to the identity principle.
Therefore we do not accept the meaning of one-electron state in the sense of physical reality.
However, the concept of one-electron state helps to consider a real interacting many-particle system as a non-interacting quasi-particle system \cite{Kaxiras} which yields the same result as real one for the collective behavior of the electron system.

We consider the Slater determinant as the mathematically optimized choice for the antisymmetrization of wavefunctions without assuming the one-electron wavefunction in the sense of the mean field approximation.
In fact, the Slater determinant does not assume the property of variable separation and can be considered as a purely mathematical selection for antisymmetric wavefunctions.
For this reason, we, without loss of generality, begin with SD.
To deduce one-electron equation, we introduced the concept of equivalent function suggestive of pseudo wavefunction.
The equivalent function enables us to obtain a single one-electron equation, keeping the antisymmetry. 
It makes us shun many hardships to be faced with when manipulating many-electron equations and leads to obtaining a one-electron equation embodying the antisymmetry property.

The method of this formalism for dealing with the electronic correlation is distinguished from the others since it is based on the uncertainty principle, Pauli's principle and phase space. 
The electronic-correlation problem initiated by Gouy and Chapman and introduced into elucidation of many-body effects by Debye and H\"{u}ckel is a non-trivial problem and a great challenge to resolving many-electron problems \cite{Chapman,Gouy,Debye}.
The electronic correlation can be understood to have classical plus quantum cause.
Our approach for considering the electronic correlation focuses on its quantal cause.
The present approach allows us to avoid assuming an additional correlation term in the wave equation for many-electron systems.
The phase norm enables us to explain the possible electron-approachable limit according to the principle of quantum mechanics, so the present approach offers the possibility of revealing  the electronic correlation depending upon its quantum nature. 
For the KS approach, finding the exact exchange-correlation functional is accompanied by painful processes of modeling. 
The different understandings of the electronic exchange and correlation in DFT have yielded a great diversity of calculation approaches \cite{Becke1,Perdew1,Gusarov,Thiele,Weber,Guidez,Mark,Higuchi2,Jarborg,Pitarke,Hanas,Sala,Tran,Arbuznikov,White,Francesco,Lein,Kurth}.
Although there has been considerable progress in investigating the exchange and correlation, it still remains as a core problem for first-principles calculations. 
Such a situation shows that there is not seen the prospect of finding the final solution to the exchange and correlation problem by manipulating within DFT.
The present method circumvents the obstacles of DFT and HF theory, by employing the equivalent function and phase norm.
It is concluded that the derived equation amounts to a one-electron equation which does not include any additional term pertaining to the exchange and correlation, but takes into consideration the exchange and correlation effects sufficiently in an explicit way.

The equation addresses electronic states of a many-electron system with single-electron orbitals, thereby yielding eigenfunctions and energy eigenvalues of the many-electron system.
Obviously, it is an equation for many-electron systems taking on the form of a single equation similar to the Kohn-Sham equation.
We consider that one of the main goals of the ab initio calculation for many-electron systems is to obtain a one-electron equation with the explicit expression of exchange and correlation depending upon  the nature of identical-particle system. 
Promising for HF approach is that in spite of   neglecting the electronic correlation and not being framed as a single equation, it provides the explicit representation of exchange with the help of antisymmetric wavefunction for a many-electron system.
On the other hand, promising for the Kohn-Sham approach is that it takes on the form of a single electron equation including exchange and correlation effects, though in a factitious way.
It is desirable to conceive a one-electron equation representing both the exchange and correlation in an explicit way, and in addition, excluding the self-interaction.
The present one-electron equation includes the interaction between electrons in an external potential.  
The absence of interaction between particles leads to the simplest model, namely IPM for which the Hamiltonian operator contains only one-body term.
Even if this model addresses the simplest case, it would formulate the fundamental framework at which the research on many-particle systems should arrive, provided that one could devise the techniques for representing the interaction between particles in the form of one-body potential in the presence of an external field. 
Our one-electron equation resolves the problem of how to describe in the one-body context the exchange and correlation effects inherent in many-electron systems.

This formalism is possible to be easily extended to excited state problems, since it has the advantage of one-electron equation and explicit exchange-correlation representation.
Originally, the excited-state problem is not distinguished from the ground-state problem because they are both stationary-state problem.
The difference between two problems is that the ground state corresponds to the lowest extreme value of a total energy functional, while the excited states relate to the rest.
Therefore we have shown that it is possible and effective to elucidate electronic states of many-electron systems in a unified and self-consistent way using the one-electron equation, irrespective of  ground or excited states. 

\section{\label{sec:Conclusion}Conclusion}

We have obtained a one-electron equation which includes exchange and correlation effects in an explicit way, and enables us to resolve the ground and excited states problem in a unified way.
The goal of obtaining a one-electron equation of the desired characteristics has been achieved by introducing the equivalent function and  the correlation-hole function in terms of the phase norm. 

This approach can be considered as an integration of the advantages of several approaches to the calculation for electronic states.
Unlike HF equation, the obtained equation is a one-electron equation which fulfills the requirement that electronic states of many-electron systems should be determined by a single eigenvalue equation since identical particles are not distinguished from one another.
An advantage over HF theory using the complicated methods with combination of SDs to explain the electronic correlation is that this method uses the explicit and direct expression of the electronic correlation depending on the quantum nature.
Unlike the KS equation that adopts several formal ways to take into consideration exchange and correlation effects, this equation does not include the additional exchange-correlation term incompatible to quantum nature.
In conclusion, the derived equation fulfills the primary requirements for the ab initio calculation concerning the self-interaction, exchange-correlation and one-electron equation excluding any empirical methods.

We have shown that the present one-electron equation can be used to describe excited states as well as ground states in an identical way.
The present method of calculation for excited states can be extended to DFT and HF approach, since they also use one-electron wavefunctions.

We believe that thanks to methodological characteristics, the present work can contribute to exploring a new avenue towards the decisive solution to the exchange-correlation and the simple but precise calculation for electronic stationary states including excited states by means of a unified method. 

\section*{\label{ack}Acknowledgments}
This work was supported partially from the Committee of Education, Democratic People's Republic of Korea, under the project entitled ``Ab Initio Calculation: Exchange-Correlation Effects and Excited States''.
 We thank Profs. Chol-Jun Yu and Hak-Chol Pak from Kim Il Sung University for useful advice and help. 
We are grateful to Prof. Il-Yong Kang from Kimchaek University of Technology for helpful information and comment.


\begin{thebibliography}{37}
\bibitem{Hartree} D. R. Hartree, {\it Proc. Cambridge Phil. Soc.} {\bf 24},  89--110 (1928)
\bibitem{Slater} J. C. Slater, {\it Phys. Rev.} {\bf 32}, 339-348 (1928)
\bibitem{Kohanoff} J. Kohanoff, {\it Electronic structure calculations for solids and molecules: theory and computational methods}, Cambridge university press, Cambridge, 2006
\bibitem{Fock} V. Fock, {\it Z. Phys.} {\bf 61}, 126--148 (1930) 
\bibitem{Lipparini} E. Lipparini, {\it Modern Many-Particle Physics: Atomic Gases, Quantum Dots and Quantum Fluids}, World Scientific, Newjersey, London, 2003
\bibitem{Jensen}  F. Jensen, {\it Introduction to computational chemistry, Wiley}, 1999
\bibitem{Davidson}  E. R. Davidson, {\it J. Compt. Phys.} {\bf 17},  87--94 (1975)
\bibitem{Hohenberg} P. Hohenberg, W. Kohn, {Phys. Rev. B} {\bf 136}, 864--867 (1964)
\bibitem{Kohn} W. Kohn, L. Sham, {\it Phys. Rev.} A {\bf 140}, 1133--1138 (1965)
\bibitem{McGUIRE} J. H. McGUIRE, {\it Electron correlation dynamics in atomic collisions}, Cambridge university press, Cambridge, 1997
\bibitem{March} N. M. March, {\it Electron correlation in the solid state}, Imperial College Press, London, 1999
\bibitem{Karasiev}  V. V. Karasiev, T. Sjostrom, J. Dufty, S. B. Trickey, {\it Phys. Rev. Lett.} {\bf 112}, 076403 (2014)
\bibitem{Cafiero} M. Cafiero, {\it Chem. Phys. Lett.} {\bf 418}, 126--131 (2006)
\bibitem{Perdew5}  J. P. Perdew, K. Burke, M. Ernzerhof, {\it Phys. Rev. Lett.} {\bf 77}, 3865 (1996) 
\bibitem{Boese1}  A. D. Boese, N. L. Doltsinis \textit{et al}, {\it J. Chem. Phys.} Vol. {\bf 112}, No. {\bf 4}, 1670--1678 (2000) 
\bibitem{Boese2} A. D. Boese, N. C. Handy, {\it J. Chem. Phys.} {\bf 88(4)}, 5497--5503 (2001)
\bibitem{Constantin}  L. A. Constantin, E. Fabiano, F. D. Sala, {\it J. Chem. Theory Compt.} {\bf 9}, 2256--2263 (2013) 
\bibitem{Becke1}  A. D. Becke, {\it J. Chem. Phys.} {\bf 84 (8)}, 15 April,  4524--4529 (1986)
\bibitem{Becke2}  A. D. Becke, {\it Phys. Rev.} A Vol. {\bf 38} No. {\bf 6}, 3098--3100 (1988) 
\bibitem{Becke3}  A. D. Becke, {\it J. Chem. Phys.} {\bf 107 (20)}, {\bf 22} November, 8554--8560(1997)
\bibitem{Sharma}  S. Sharma, J. K. Dewhurst,  A. Sanna, {\it Phys. Rev. Lett.} {\bf 107}, 186401 (2011) 
\bibitem{Perdew1} J. P. Perdew, W. Yue, {\it Phys. Rev.} B  Vol {\bf 33}, No. {\bf 12},  8800--8802 (1986)
\bibitem{Perdew3}  J. P. Perdew, W. Yue, {\it Phys. Rev.} B Vol. {\bf 45}, No. {\bf 23}, 13244-13249 (1992)
\bibitem{Gusarov}  S. Gusarov, P. Malmqvist, R. Lindh, B. O. Roos, {\it Theor. Chem. Acc.} (Theoretical Chemistry Accounts) {\bf 12}, 84--94 (2004) 
\bibitem{Gonzalez} L. Gonz\'{a}lez, D. Escudero, L. Serrano-Andr\'{e}s, {\it ChemPhysChem} {\bf 13}, 28--51 (2012)
\bibitem{Stratmann} R. E. Stratmann, G. E. Scuseria, M. J. Frisch, {\it J. Chem. Phys.} {\bf 109}, 8218 (1998)
\bibitem{Kilina} S. Kilina, E. Badaeva, A. Piryatinski, S. Tretiak, A. Saxena, A. R. Bishop, {\it Phys. Chem. Chem. Phys.} {\bf 11}, 4113 (2009)
\bibitem{Gross} E. K. U. Gross, J. F. Dobson, M. Petersilka, {\it Top. Curr. Chem.} {\bf 181}, 81 (1996)
\bibitem{Jamorski} C. Jamorski, M. E. Casida, D. R. Salahub, {\it J. Chem. Phys.} {\bf 104}, 5134 (1996)
\bibitem{Dreuw1} A. Dreuw, M. Head-Gordon, {\it Chem. Rev.} {\bf 105}, 4009 (2005)
\bibitem{Dreuw2} A. Dreuw, M. Head-Gordon, {\it J. Am. Chem. Soc.} {\bf 126}, 4007 (2004)
\bibitem{Roos} B. O. Roos, M. Merch\'{a}n, R. McDiarmid, X. Xing, {\it J. Am. Chem. Soc.} {\bf 116}, 5927 (1994).
\bibitem{Nooijen1} M. Nooijen, R. J. Bartlett, {\it J. Chem. Phys.} {\bf 106}, 6449 (1997)
\bibitem{Nooijen2} M. Nooijen, R. J. Bartlett, {\it J. Chem. Phys.} {\bf 106},  6441 (1997)
\bibitem{Koch} H. Koch, O. Christiansen, P. Jorgensen, J. Olsen, {\it Chem. Phys. Lett.} {\bf 244}, 75 (1995)
\bibitem{Falden} H. H. Falden, K. R. Falster-Hansen, K. L. Bak, S. Rettrup, S. P. A. Sauer, {\it J. Phys. Chem.} A {\bf 113}, 11995 (2009)
\bibitem{Oddershede} J. Oddershede, {\it Adv. Chem. Phys.} {\bf 69},  201 (1987)
\bibitem{Packer} M. J. Packer, E. K. Dalskov, T. Enevoldsen, H. J. A. Jensen, J. Oddershede, {\it J. Chem. Phys.} {\bf 105}, 5886 (1996).
\bibitem{Kaxiras} E. Kaxiras, {\it Atomic and Electronic
Structure of Solids}, Cambridge university press, Cambridge, 2003.
\bibitem{Chapman} D. L. Chapman, {\it Philos. Mag. Lett.} {\bf 25}, 475--481 (1913)
\bibitem{Gouy} G. Gouy, {\it J. Phys. Theor. Apple.} {\bf 9}, 455--468 (1910)
\bibitem{Debye} P. Debye, E. H\"{u}ckel, {\it Phys. Z.} {\bf 24}, 185--206 (1923)
\bibitem{Thiele}  M. Thiele, S. K\"{u}mmel, {\it Phys. Rev. Lett.} {\bf 112}, 083001 (2014). 
\bibitem{Weber}  A. Weber, F. Astorga, {\it Int. J. Mod. Phys.} A Vol. {\bf 29}, No. {\bf 5}, 1450018 (2014) 
\bibitem{Guidez}  E. B. Guidez, M. S. Gordon, {\it J. Phys. Chem.}  A {\bf 119}, 2161--2168 (2015)
\bibitem{Mark}  M. R. Mark, D. Whitenack, A. Wasserman, {\it Chem. Phys. Lett.} {\bf 558}, 15--19 (2013). 
\bibitem{Higuchi2}  K. Higuchi, M. Higuchi, {\it Physica} B {\bf 312-313}, 534-536 (2002) 
\bibitem{Jarborg}  T. Jarborg, {\it Phys. Lett.} A {\bf 260}, 395--399 (1999)
\bibitem{Pitarke}   J. M. Pitarke,  L. A. Constantin, J. P. Perdew, {\it Phys. Rev} B {\bf 74}, 045121 (2006) 
\bibitem{Hanas}  M. Hanas, M. Markowski, {\it Comp. Theo. Chem.} {\bf 1060}, 52--57 (2015) 
\bibitem{Sala}  F. D. Sala, A. G\"{o}rling, {\it Phys. Rev. Lett.} Vol. {\bf 89}, 033003 (2002) 
\bibitem{Tran}  F. Tran, P. Blaha, K. Schwarz, {\it J. Phys. Condens. Matter} {\bf 19}, 196208 (2007). 
\bibitem{Arbuznikov}  A. V. Arbuznikov, M. Kaupp, {\it Chem. Phys. Lett.} {\bf 440}, 160--168 (2007) 
\bibitem{White}  J. A. White, D. M. Bird, {\it Phys. Rev.} B  Vol. {\bf 50}, No. {\bf 7}, 4954--4957 (1994) 
\bibitem{Francesco}  F. Muniz-Miranda, M. C. Menziani, A. Pedone, {\it J. Phys. Chem.} C {\bf 118}, 7532--7544 (2014)
\bibitem{Lein}  M. Lein, S. K\"{u}mmel, {\it Phys. Rev. Lett.} {\bf 94}, 143003 (2005) 
\bibitem{Kurth}  S. Kurth, {\it J. Mol. Struc.} (Theochem), 189-194 (2000)


\end{thebibliography}
\end{document}